\documentclass[12pt]{article}
\usepackage{cite}
\newcommand{\be}{\begin{equation}}
\newcommand{\bea}{\begin{eqnarray}}

\newcommand{\eea}{\end{eqnarray}}
\newcommand{\ba}{\begin{array}}
\newcommand{\ea}{\end{array}}
\newcommand{\ee}{\end{equation}}

\expandafter\ifx\csname mathbbm\endcsname\relax

\else

\fi
\begin{document}
\input epsf
\begin{titlepage}
\hfill
\vbox{
    \halign{#\hfil         \cr
           IPM/P-2006/013 \cr
           TIFR/TH/06-06 \cr 
           } % end of \halign
      }  % end of \vbox
\vspace*{20mm}
\begin{center}
{\Large {\bf Partition function of non-supersymmetric\\black holes in the supergravity limit}}\\ 

\vspace*{15mm} \vspace*{1mm} {Shahrokh Parvizi$^{a}$\footnote{parvizi@theory.ipm.ac.ir} and Alireza Tavanfar$^{a,b}$
\footnote{art@ipm.ir}}
 \\
\vspace*{1cm}

{$^a$ Institute for Studies in Theoretical Physics
and Mathematics (IPM)\\
P.O. Box 19395-5531, Tehran, Iran \\
\vspace*{3mm}
$^b$ Tata Institute of Fundamental Research \\
Homi Bhabha Road, Mumbai, 400 005, INDIA\\}

%%\maketitle
\end{center}

\begin{abstract}
In this note, we propose the free energy of general \emph{non-supersymmetric} black hole attractors arising in type IIA(B) superstrings on 3-fold Calabi-Yau, in the supergravity limit. This, by definition, differs from its counterpart BPS free energy by a factor of $4\;$. Correspondingly, a mixed ensemble for these black holes is proposed. 
\end{abstract}

\end{titlepage}
\section{Introduction}
Non-supersymmetric black holes are interesting phenomenologically but hard to handle theoretically. Recently due to the discovery of \emph{non-supersymmetric attractors}  \cite{Sen:2005wa,Goldstein:2005hq,Kallosh:2005ax}, new perspectives welcome exploring such black holes. For earlier works hinting to the fact that BPS condition is not necessary for extremal attractors see \cite{Gibbons:1996af,Ferrara:1997tw}. Works in this field include  \cite{Sen:2005iz,Tripathy:2005qp,Giryavets:2005nf,Prester:2005qs,Goldstein:2005rr,Alishahiha:2006ke,Sinha:2006yy,Kallosh:2006bt,Chandrasekhar:2006kx,Bellucci:2006ew,Kallosh:2006bx}. Such attractors can be black hole solutions to either a non-supersymmetric theory of gravity coupled to the gauge fields and moduli fields \cite{Goldstein:2005hq,Chandrasekhar:2006kx}, or non-BPS solutions to a supersymmetric theory \cite{Tripathy:2005qp,Kallosh:2006bt}. In this note we are interested in some examples of the second class in the in the large charge and large volume Calabi-Yau limit. More precisely, following the studies of \cite{Tripathy:2005qp,Kallosh:2006bt}, we consider non-supersymmetric black hole attractors which arise in $CY_3$ compactifications of type IIA(B) superstrings, where the effective theory enjoys $\mathcal{N}=2$ supersymmetry and (the F-term sector) admits the well-developed structure of symplectic geometry. In the supergravity limit the prepotential is given by, \bea \label{prepoten0} F(X)\;=D_{ijk}\frac{X^i X^j X^k}{X^0}\;\;\;;\;\;\;i,j,k = 1..n\;\;\;.\eea To be an attractor, a non-supersymmetric black hole in this theory should be \emph{extremal} \cite{Goldstein:2005hq}, and should \emph{not} carry the $D6$-brane charge \cite{Tripathy:2005qp}. This note concerns attractors and so these two requirements are assumed to be the case from now on. In \cite{Tripathy:2005qp,Kallosh:2006bt} the corresponding attractor solution was found. Then, one question which is natural to be addressed is evaluating the \emph{partition function} of these black holes. This is the subject of this note. Denoting the degeneracy of black hole states with $ \rm{d} (p,q) = \rm{d} [ D(p)M ]$, given the black hole charge-multiplet $(p^i,q_i,q_0)$, we shall identify the mixed partition function of the non-susy black hole as, 
\bea\label{result}Z_{_{ns.BH}}\equiv\sum_{M} \rm{d} [ D(p)M ]\; e^{-M\;\Im [C X^0]}\;=\;\rm{e}^{\frac{\pi}{2}\frac{D(p)}{\Im [C X^0]}}\;\;\;\eea 
where $M = -2 \pi \hat{q}_0$ and \bea\label{conven}\hat{q}_0 \equiv q_0+\frac{1}{12}D^{ij}(p)q_i q_j\;;\;  D^{ij}(p)D_{jk}(p) = \delta^{i}_{k}\;;\;D_{ij}(p) \equiv D_{ijk} p^k\;;\;D(p) \equiv D_{ijk} p^i p^j p^k .\eea The way we find this result is similar to the analysis of \cite{Ooguri:2004zv}, in essence. We first rewrite the general attractor equations of \cite{Kallosh:2006bt} in a slightly different form, which is more similar to the language of \cite{Ooguri:2004zv}, as follows,
\bea \label{allpC} p^I\;=\;\Re\;[\;C\;X^I\;-\;G^{j\bar{l}}\;\bar{\partial}_{\bar{l}}\bar{C}\; \nabla_{j}X^I\;\;]\;\;\;\eea \bea \label{allqC} q_I\;=\;\Re\;[\;C\;F_I\;-\;G^{j\bar{l}}\;\bar{\partial}_{\bar{l}}\bar{C}\; \nabla_{j}F_I\;\;]\;\;\;\eea
\bea\label{CCC}2\;C\;\partial_i \bar{C}\;-\;\mathbf{i}\;C_{ijk}\;G^{j\bar{n}}\;G^{k\bar{l}}\;\bar{\partial}_{\bar{n}}C\;\bar{\partial}_{\bar{l}}C\;=\;0\;\;\;. \eea
At the next step we reformulate the original OSV ensemble, which is for BPS black holes, in a \emph{reduced} but equivalent form. Then, motivated by this reformulation, we show that for the non-supersymmetric solution of (\ref{allpC}-\ref{CCC}), the black hole entropy is related via a Legendre transformation to the \emph{free energy}, \bea\label{freeE}\mathcal{F}_{_{ns.BH}}\;=\;\frac{\pi}{2}\frac{D(p)}{\Im [C X^0]}\;\;\;.\eea The Legendre transformation reads as,
\bea\label{LT}S_{_{ns.BH}}(p,q) = \mathcal{F}_{_{ns.BH}} - \Im(C X^0)\; \frac{\partial \mathcal{F}_{_{ns.BH}}}{\partial \Im(C X^0)}\eea with,\bea\label{q0}\hat{q}_0\;=\;\frac{1}{2 \pi}\;\frac{\partial\; \mathcal{F}_{_{ns.BH}}}{\partial\; \Im(C X^0)}\;\;\;.\eea
The free energy (\ref{freeE}) then supports for the mixed partition function given by (\ref{result}). 
Here there are two points which we would like to highlight: \\$1.\;$  In the technical sense, the sum in (\ref{result}) should be interpreted carefully. The problem is that $\rm{d}(p,q)$ as
derived by the inverse Laplace transformation corresponding to (\ref{result}), does not necessarily respect the expected symplectic symmetry. The same problem is the case for the mixed OSV ensemble for which there are two ways to restore this symmetry. One way is incorporating the sum with an appropriate measure, for example \cite{Dabholkar:2005by,Dabholkar:2005dt,Shih:2005he,LopesCardoso:2006bg} suggest such measures for the case of OSV ensemble. An equivalent way is based on the approach of
\cite{Parvizi:2005aa}. One uses the fact that the Legendre transformation which bridges between the free energy (\ref{freeE}) and the black hole entropy is preserved up to a rescaling of the `potential'      $\Im(C X^0)$. Then applying the symplectic symmetry as a \emph{constraint} on $\rm{d}(p,q)$, a proper charge-dependent rescaling of $\Im(C X^0)$ can be chosen to redefine the ensemble. \\ $2.\;$Unlike the main stream of \cite{Ooguri:2004zv} which relates the free energy of 4-dimensional BPS black holes with
the free energy of topological strings, we do \emph{not} connect the non-BPS partition function (\ref{result}) with topological strings. That is although using  (\ref{newnsdeduc}), (\ref{newfree}), (\ref{nshatfree}) and (\ref{ff}) one easily gets an OSV-like relation at the supergravity limit, such formal connection faces some serious difficulties when treated as a meaningful relation\footnote{We are thankful to Cumrun Vafa for discussion in this regard.}.\\
The outline of this work is as follows. In section $2$, we review the derivation of the general attractors,  rewrite them in the form of (\ref{allpC}-\ref{CCC}) and recover the results of \cite{Tripathy:2005qp,Kallosh:2006bt}. In Section $3$, through $3$ subsections, we first give a brief recap of the OSV proposal, then introduce an equivalent reduced form of the OSV ensemble and finally find the partition function of the non-supersymmetric attractors in the supergravity limit.
We end this note with a short list of some open problems.
%%%%%%%%%%%%%%%%%%%%%%%%%%%%%%%%%%%%%%%%%%%%%%%%%%%%%%%%%%%%%%%%%%%%%%%%%%%%%%%%%%%%%%%%%%%%%%%%%%%%%%%%%
\section{Attractor equations} It is easier to tell the story in the IIB language. Mirror symmetry then takes us to the IIA dual. Let us begin with an arbitrary 3-cycle $\Gamma$ living in the 3-fold Calabi-Yau $\mathcal{X}$ on which one compactifies the theory. In a symplectic basis $(A^I,B_I)$, with $I$ running from $0$ to $n+1$, the unique decomposition of $\Gamma$ reads as, $$\Gamma\;=\;q_I A^I - p^I B_I\;\;\;. $$ The Poincare dual of this cycle is a 3-form $G_3$, called the 3-form flux, which in the dual basis $(\alpha_I,\beta^I)$ takes the form, \bea \label{GVW} G_3\;=\;p^I \alpha_I - q_I \beta^I\;\;\;. \eea  In this flux background the GVW superpotential \cite{GVWsup} is defined as,$$ W \equiv \int G_3 \wedge \Omega $$ with $\Omega$ being the holomorphic 3-form of $\mathcal{X}$. In terms of the \emph{symplectic period variables}, $$X^I \equiv \int_{A^I} \Omega\;\;\;\;;\;\;\; F_I \equiv \int_{B^I} \Omega = \partial_I F(X) $$ the superpotential reads as, $$W\;=\;q_I X^I\;-\;p^I F_I\;\;\;. $$ Let further parametrize the complex structure moduli of $X$ by $t^i \equiv \frac{X^i}{X^0}$, $i\;=\;0...n\;$, and $\nabla_i$ stand for (the components of) the K\"ahler covariant derivative,  $$\nabla_i X^J \equiv \partial_i X^J + (\partial_i K) X^J\;\;\;;\;\;\; \bar{\nabla}_{\bar{i}} X^J \equiv 0\;\;\; $$ where the K\"ahler potential is given by, \bea \label{Kahler}   K = - \rm{ln} [\;\mathbf{i}\;(\bar{X}^J F_J - X^J \bar{F}_J)\;]\;\;\;.\eea As explored in \cite{Ferrara:1995ih,Strominger:1996kf,Ferrara:1996dd,Ferrara:1996um} for BPS black holes, and recently (in references given in the introduction) for non-BPS black holes, an extremal solution generally enjoys the \emph{attractor} mechanism, according to which the moduli fields with free asymptotic values at infinity take some fixed values on the horizon, as determined by the black hole charges. More precisely, these attractor values are uniquely determined by \emph{minimizing} an effective potential. In a theory with $\mathcal{N}=2$ supersmmetry, this effective potential is given in terms of the GVW superpotential as, \bea \label{eff} V_{eff} \equiv \rm{e}^K\;(\;|W|^2 + |\nabla_i W|^2\;)\;\;\;. \eea The black hole is \emph{BPS} if and only if, \bea \label{BPS}\nabla_i W\;=\;0\;\;\;\eea which is the trivial (minimum) solution of \bea \label{effeq} \nabla_i V_{eff}\;=\;\partial_i V_{eff}\;=0\;\;\;. \eea Non-supersymmetric black holes correspond to non-trivial (minimum) solutions of (\ref{effeq}). Recently \cite{Tripathy:2005qp,Kallosh:2006bt} solved these equations in the supergravity limit which corresponds to (\ref{prepoten0}). They showed that for (\ref{prepoten0}), equations (\ref{effeq}) admit either a BPS solution if $\hat{q}_0 D(p) \geq 0 $ or a non-supersymmetric solution if  $\hat{q}_0 D(p) \leq 0 $, with the convention (\ref{conven}). To get this result, \cite{Tripathy:2005qp} simply plugs (\ref{GVW}) back into (\ref{effeq}). \cite{Kallosh:2006bt} however develops a different approach which we report here briefly. Given $\Omega$, the Hodge-decomposition, $$H^{3}(X) = H^{3,0}\; \oplus\; H^{2,1}\; \oplus\; H^{1,2} \;\oplus\; H^{0,3}$$ admits the basis, $$\Omega\;\oplus\;\nabla_i \Omega\;\oplus\;\bar{\nabla}_{\bar{i}} \bar{\Omega}\;\oplus\;\bar{\Omega}\;\;\;.$$ In this basis, a real 3-form like $G_3$ admits the general expansion, \bea\label{gexpan} G_3 \;=\; \mathbf{i}\;\rm{e}^{\frac{K}{2}}\;[\;\bar{x}\; \bar{\Omega} +\; \bar{x}^{\bar{j}}\;\bar{\nabla}_{\bar{j}}\bar{\Omega}\; -\; c.c.\;] \;\;\;. \eea 
Now taking $\int \Omega \wedge$ of both sides of (\ref{gexpan}) and using the identity,
$$\mathbf{i} \int \Omega \wedge \bar{\Omega}\;=\;\rm{e}^{-K} $$ one gets $\bar{x}=-\;Z$, with $Z \; \equiv \rm{e}^{K/2} W$ defined as the (covariantly holomorphic) \emph{central charge}. Similarly applying       $\int \nabla_l \Omega \wedge$ to both sides of (\ref{gexpan}) and using,
$$\mathbf{i}\;\int \bar{\nabla}_{\bar{j}}\bar{\Omega} \wedge \nabla_{l}{\Omega}\;=\;\rm{e}^{-K}\;G_{l\bar{j}}$$ one gets $\bar{x}^{\bar{j}}\;=\;G^{l\bar{j}}\;\nabla_{l}Z$. Thus the general expansion of the 3-form flux reads as, \bea \label{gexpanIm} G_3\;=\;2\;\rm{e}^{K}\;\Im\;[\;W\;\bar{\Omega}\;-\; \;G^{l\bar{j}}\;\nabla_{l}W\;\bar{\nabla}_{\bar{j}}\bar{\Omega}\;]\;\;\;.\eea Subsequently integrating both sides of $\ref{gexpanIm}$ over the cycles $(A^I,B_I)$ yields,\bea \label{pIm} p^I\;=\;2\;\rm{e}^{K}\;\Im\;[\;W\;\bar{X}^I\;-\;G^{l\bar{j}}\;\nabla_{l}W\;\bar{\nabla}_{\bar{j}}\bar{X}^I\;]\;\;\;\eea and \bea \label{qIm} q_I\;=\;2\;\rm{e}^{K}\;\Im\;[\;W\;\bar{F}_I\;-\;G^{l\bar{j}}\;\nabla_{l}W\;\bar{\nabla}_{\bar{j}}\bar{F}_I\;]\;\;\;.\eea To go further we use the identities,
$$\nabla_{i}\bar{W}\;=\;0$$ $$\nabla_{i}\bar{\nabla}_{\bar{j}}W\;=\;G_{i\bar{j}}\;W$$             $$D_{i}\nabla_{j}W\;=\;\mathbf{i}\;C_{ijk}\;G^{k \bar{l}}\;\bar{\nabla}_{\bar{i}}\nabla_{l}W $$
where the real symmetric coefficients $C_{ijk}$ are given as,
\bea\label{cijk}C_{ijk} = \rm{e}^K\;(\nabla_{i}X^L\;\nabla_{j}X^M\;\nabla_{k}X^N\;C_{LMN})\;\;\;;\;\;\;    C_{LMN}\;\equiv\;\partial_{L}\partial_{N}\partial_{M}\;F(X)\eea and
the operation $D_i$ involves the Christoffel connection as well as the K\"ahler connection, i.e. schematically, $D_{i}\;\equiv\;\nabla_{i}\;+\;\Gamma_{i}$. Now computing $ D_i V \;=\; \nabla_i V $, via (\ref{eff}), the equation (\ref{effeq}) takes the form,
\bea \label{effeqW} 2 \nabla_i W\; \bar{W}\;+\;\mathbf{i}\;C_{ijk}G^{j\bar{m}}G^{k\bar{n}}        \;\bar{\nabla}_{\bar{m}}\bar{W}\; \bar{\nabla}_{\bar{l}}\bar{W}\;=\;0\;\;\;. \eea Finally
\cite{Kallosh:2006bt} reads $\nabla_i W$ from (\ref{effeqW}), substitutes it into the equations (\ref{pIm}), (\ref{qIm}) and solves the resulted equations to get the attractor values of the moduli, the result of course matches with that of \cite{Tripathy:2005qp}.\\ To cook up our cake, we recast the equations (\ref{pIm}), (\ref{qIm}) and (\ref{effeqW}) in a new form which is closer to the original OSV language. Let's define an object $C$ as, \bea\label{Cdef}C\;\equiv\;2\;\mathbf{i}\;\rm{e}^K\;\bar{W}\;\;\;\eea for  which the following \emph{identities} hold,
$$\nabla_i C\;=\;0\;\;\;;\;\;\;\bar{\nabla}_{\bar{i}}C\;=\;\bar{\partial}_{\bar{i}}C\;\;\;.$$
It is worth to note that the combinations $CX^I$ and $CF_I$ are invariant under the \emph{K\"ahler gauge} transformations,                                                                            $ K(t,\bar{t})\;\rightarrow\;K(t,\bar{t})-f(t)-\bar{f}(\bar{t}) $.\\
In terms of (\ref{Cdef}), the equations (\ref{pIm}), (\ref{qIm}) and (\ref{effeqW}) take the following final forms, 
\bea\label{CC}2\;C\;\partial_i \bar{C}\;-\;\mathbf{i}\;C_{ijk}\;G^{j\bar{n}}\;G^{k\bar{l}}\;\bar{\partial}_{\bar{n}}C\;\bar{\partial}_{\bar{l}}C\;=\;0\;\;\; \eea
\bea \label{pC} p^I\;=\;\Re\;[\;C\;X^I\;-\;G^{j\bar{l}}\;\bar{\partial}_{\bar{l}}\bar{C}\; \nabla_{j}X^I\;\;]\;\;\;\eea \bea \label{qC} q_I\;=\;\Re\;[\;C\;F_I\;-\;G^{j\bar{l}}\;\bar{\partial}_{\bar{l}}\bar{C}\; \nabla_{j}F_I\;\;]\;\;\;.\eea  
We call (\ref{CC})  the `` C-equations " and (\ref{pC},\ref{qC}) the `` charge equations ". They, \emph{as a whole}, form the most general \emph{attractor equations} in the $\mathcal{N}=2$ effective theory of $IIB(A)$ on $CY_3$. Having these equations at hand, one can safely forget the original definition of $C$, because they not only determine the attractor value of $C$ but also imply (\ref{Cdef}) as consequent relation. In fact to solve the equations (\ref{pC}), (\ref{qC}) and (\ref{CC}), the best way is treating them as a coupled system of $(4n+2)$ \emph{nonlinear algebraic equations} governing $(4n+2)$ unknown variables. These variables are $(t^i, C ; \bar{\partial}_{\bar{i}} C)$ together with their complex conjugates. Although one is only interested in determining the attractor values of $t^i$, since the equations (\ref{CC}-\ref{qC}) are strongly coupled, one needs to determine all of these variables.\\ The trivial solution of (\ref{CC}) is given simply as, \bea\label{BPSat}\partial_i \bar{C}\;=\;\bar{\partial}_{\bar{i}} C\;=\;0\eea which result
at the well known \emph{BPS attractor} equations \cite{Ooguri:2004zv},                                  \bea\label{pBPS}p^I = \Re\;[\;C X^I\;]\eea    \bea\label{qBPS}q_{I} = \Re\;[\;C F_{I}\;]\;\;\;.\eea  \\ As a warm up for the goal of this paper, we end this section with re-deriving the results of \cite{Tripathy:2005qp,Kallosh:2006bt} using the above set of equations. We set $p^0=0$ which is needed for the non-supersymmetric attractor and also avoids various conceptual and technical difficulties in the BPS case. Further without loss of generality, we can safely restrict the case to the simple STU model defined as, \bea \label{STU}F\;=\;\frac{X^{1}X^{2}X^{3}}{X^0}\;\;\;;\;\;\;(S\;,T\;,U)\;\equiv\;(t^1\;,t^2\;,t^3)\;\;\;\eea and set $q_i=0$. As soon as one gets the final results, the dictionary to translate them to the general case of (\ref{prepoten0}) with $q_i \neq 0 $ is clear and simple, as we shall finally do. Working in the $X^0 = 1$ gauge, from (\ref{STU}), (\ref{Kahler}) and (\ref{cijk}) one finds $$K = -\rm{ln}\;(-8 \; S_2 T_2 U_2)\;\;\;;\;\;\;C_{ijk} = \frac{-1}{8 \; S_2 T_2 U_2}\; |\epsilon_{ijk}|\;\;\;. $$ where $(S_2,T_2,U_2)$ stand for the imaginary parts of the moduli. It is easy to observe that the charge equations match with the conditions $p^0=q_{a}=0$, if the moduli and $C$ take \emph{pure imaginary values}, $$t^{j} = \mathbf{i}\;t_2^{j}\;;\;\forall j$$ $$C(S,T,U) = \mathbf{i}\; C_2(S_2,T_2,U_2)\;\;\;\;.$$ Then denoting,
$$ C_{2,j} = \frac{\partial C_2}{\partial t_2^j} $$
the nontrivial charge and C-equations take the form,
\bea\label{eq1}C_2 C_{2,1} - \frac{T_2 U_2}{S_2}\; C_{2,2} C_{2,3} = 0\eea
\bea\label{eq2}C_2 C_{2,2} - \frac{U_2 S_2}{T_2}\; C_{2,3} C_{2,1} = 0\eea
\bea\label{eq3}C_2 C_{2,3} - \frac{S_2 T_2}{U_2}\; C_{2,1} C_{2,2} = 0\eea
\bea\label{eq4}\frac{p^1}{S_2} = S_2 C_{2,1} - T_2 C_{2,2} - U_2 C_{2,3} - C_2\eea
\bea\label{eq5}\frac{p^2}{T_2} = - S_2 C_{2,1} + T_2 C_{2,2} - U_2 C_{2,3} - C_2\eea
\bea\label{eq6}\frac{p^3}{U_2} = - S_2 C_{2,1} - T_2 C_{2,2} + U_2 C_{2,3} - C_2\eea       \bea\label{eq7}\frac{q_{0}}{S_2 T_2 U_2} = S_2 C_{2,1} + T_2 C_{2,2} + U_2 C_{2,3} - C_2 \eea
It turns out that if one varies the charges, the moduli space of solutions to the above set of equations is topologically disconnected. It is the union of two disjoint sectors, ` susy sector ' and ` non-susy sector ', which are characterized by the sign of $q_0 \mathcal{D}$, with $\mathcal{D} \equiv p^1 p^2 p^3$.\\ Given the charge multiplet $(p^1,p^2,p^3,q_0)$, the susy solution of (\ref{eq1}-\ref{eq7}) reads as,\bea\label{susyAt}
(\;C_{2,j} = 0\;\;\;;\;\;\;t_2^j = -\frac{p^j}{C_2}\;\;\;;\;\;\;C_2^2 = \frac{\mathcal{D}}{q_0}\;)\eea
while the non-susy solution is given by, 
\bea\label{nonsusyAt} (\;C_{2,j} = \frac{C_2}{t_2^j}\;\;\;;\;\;\;t_2^j = -\frac{p^j}{2 C_2}\;\;\;;\;\;\;    C_2^2 = -\frac{\mathcal{D}}{4 q_0}\;)\;\;\;.\eea To get the above attractor solutions, the easiest way is
treating (\ref{eq1}-\ref{eq7}) as a coupled system of $7$ algebraic equations governing $7$ unknown variables $(t_2^j,C_2,C_{2,j})$. A different but equivalent way is as follows. One first solves the equations (\ref{eq1}-\ref{eq3}) as a coupled system of $3$ differential equations. Apart from the susy solution $C_2=const.$, this gives us a unique STU-symmetric non-susy solution as $C_2=r S_2 T_2 U_2$, with $r$ a constant value. Now substituting the non-susy solution of $C_2$ in the equations (\ref{eq4}-\ref{eq6}) one gets, $ \frac{S_2}{p1} = \frac{T_2}{p2} = \frac{U_2}{p3} = A$, with, $A^4 = \frac{-1}{2 \mathcal{D} r}$. Finally substituting all the above results in the equation (\ref{eq7}), one obtains, $A=\pm \sqrt{\frac{-q0}{\mathcal{D}}}$ and accordingly, $C_2=\pm \frac{1}{2} \sqrt{\frac{-\mathcal{D}}{q0}}$. Totally, (\ref{nonsusyAt}) is recovered.  \\We note that the attractor values of the susy and  non-susy \emph{moduli} are mapped to each other simply via an \emph{analytic continuation}, $q_0\;\rightarrow\;-q_0$. \\Now we lift the results (\ref{susyAt}) and (\ref{nonsusyAt}) up to  the general case of (\ref{prepoten0}) with nonvanishing $q_i$, simply by mapping,
\bea\label{map}q^0\;\rightarrow\;\hat{q}_0\;=\;q_0\;+\;\frac{1}{12}D^{ij}(p) q_i q_j\;\;;\;\;t^i\;\rightarrow\;\hat{t}^i\;=\;t^i-\frac{1}{6}D^{ij}(p)q_j\;\;;\;\;\mathcal{D}\;\rightarrow\;D(p)\;.\eea It is worth to highlight the observation that the value of the entropy of the extremal black hole in the supergravity limit, given by the Bekenstein-Hawking formula, has the \emph{same} value for both the BPS and non-BPS sectors \cite{Tripathy:2005qp}. It  enjoys a simple relation with the value of the effective potential at the attractor point, $$ S_{BH} = \pi V_{eff}(t^{j*}_2) $$ which according to (\ref{susyAt}), (\ref{nonsusyAt}) and (\ref{eff}) equals, \bea\label{entropy} S_{BH}\;=\;2 \pi \sqrt{|Q|} \eea with $Q\;\equiv\;\hat{q}_0 D(p)$ being the invariant charge.
Finally one more comment about the non-susy attractor is in order. Unlike the susy case where solution of (\ref{BPS}) is always a minimum of $V_{eff}$, for the non-susy solution of (\ref{effeq}) the minimality condition should be checked independently. By expanding $V_{eff}$ around the non-susy attractor point one can check that (\ref{nonsusyAt}) is indeed a minimum \cite{Tripathy:2005qp}. 
%%%%%%%%%%%%%%%%%%%%%%%%%%%%%%%%%%%%%%%%%%%%%%%%%%%%%%%%%%%%%%%%%%%%%%%%%%%%%%%%%%%%%%%%%%%%%%%%%%%%%%
\section{The mixed partition function}
%%%%%%%%%%%%%%%%%%%%%%%%%%%%%%%%%%%%%%%%%%%%%%%%%%%%%%%%%%%%
\subsection{\underline{Recap of the OSV proposal}}
Lets recap the steps that lead OSV to the mixed ensemble of \cite{Ooguri:2004zv}. They consider the BPS attractor equations (\ref{pBPS}) and (\ref{qBPS}) and first solve half of them, the magnetic-charge attractors, \bea \label{solhalf}CX^I = p^I + \mathbf{i} \frac{\phi^I}{\pi}\;\;\;.\eea Then using the fact that prepotential is homogeneous of order two, from (\ref{solhalf}) they define,\bea\label{subs} \tilde{F}(p^I,\phi^I)\equiv C^{\;2} F(X^I)_{\;\;\vert_{(\ref{solhalf})}}\;\;\;.\eea Further, ignoring  holomorphic anomaly of the prepotential at higher genus levels\footnote{For attempts to incorporate the holomorphic anomaly see \cite{Verlinde:2004ck} and the interesting proposal of \cite{LopesCardoso:2006bg}.}, they recast the entropy of BPS black holes as evaluated in \cite{LopesCardoso:1999ur,LopesCardoso:1999xn}, based on the Wald entropy formula \cite{Wald:1993nt}, in the form of a Legendre transformation from the \emph{OSV free energy},\bea\label{free} \mathcal{F}(p^I,\phi^I)\;\equiv\;- \pi \Im \;[\;\tilde{F} (p^I,\phi^I)\;]\;\;\;\eea over half of its variables as follows,
\bea\label{Legendre} S(p,q) = \mathcal{F} - \phi^I\; \frac{\partial \mathcal{F}}{\partial \phi^I}\;\;\; \eea
with, \bea\label{solsecond}q_I\;=\;-\frac{\partial \mathcal{F}}{\partial \phi^I}\;\;\;.\eea 
We note that (\ref{solsecond}) is equivalent with (\ref{qBPS}) and so completes the attractor equations.\\
The result (\ref{Legendre}-\ref{solsecond}) suggests to interprate $\mathcal{F}$ as \emph{free energy of the BPS black hole} carrying charges $(p^I,q_I)$. This corresponds to a \emph{mixed} ensemble defined as,
\bea\label{BPSensemble} Z_{BH} = \rm{e}^{\mathcal{F}} \;\equiv\; \sum_{q^I} \Omega(p,q) e^{-q_{I} \phi^{I}}\eea
with $\Omega(p,q)$ being the black hole (index) degeneracy of states.\\
Finally according to the well known relation,                                                       \bea\label{ff}                                                                                          F(C X^I, C^2 \mathcal{W}^2)\;_{\vert_{C^2 \mathcal{W}^2=256}}\;=\;\frac{2}{\mathbf{i}\pi}\;F_{top}(t^i,g_{top})\;_{\vert_{g_{top} = \pm \frac{4 \pi \mathbf{i}}{C X^0}}} \eea with $\mathcal{W}^2$ as the squared graviphoton field strength,
OSV propose a beautiful connection between 4-dimensional BPS black holes and topological strings,  \bea\label{TSBH}\mathcal{F} = F_{top} + \bar{F}_{top}\;\;\;\rightarrow\;\;\;                     Z_{BH}(p^I,\phi^I) = |Z_{top}(t^i,g_{top})|^2\;\;\;.\eea
where
\bea\label{vari}t^j = \frac{p^j + \mathbf{i}\; \phi^j / \pi}{p^0 + \mathbf{i}\; \phi^0 / \pi}\;\;\;;\;\;\;g_{top} =\pm \frac{4 \pi \mathbf{i}}{p^0 + \mathbf{i}\; \phi^0 / \pi}\;\;\;.\eea
For concrete tests of the OSV proposal see \cite{Vafa:2004qa,Dabholkar:2004yr,Aganagic:2004js} and \cite{Dabholkar:2005dt,Aganagic:2005wn}. For relevant developments and some applications see \cite{Aganagic:2005dh,Dijkgraaf:2005bp,Ooguri:2005vr,Gukov:2005iy,Gukov:2005bg} .
%%%%%%%%%%%%%%%%%%%%%%%%%%%%%%%%%%%%%%%%%%%%%%%%%%%%%%%%%%%%%%%%%%%%%%
\subsection{\underline{OSV relation in the \emph{reduced} ensemble}}
An \emph{equivalent} mixed ensemble can be obtained for BPS black holes if, instead of solving half of the attractor equations and keeping the second half untouched, one solves \emph{all but one} of them and saves the last one, say $q_0$-equaton, for the Legendre transformation which leads to the black hole free energy. This results at a \emph{minimized} version of the mixed ensemble, also supports for the relation (\ref{TSBH}) but with some other relations between the charge-potential variables and the topological variables, as compared to (\ref{vari}).
\\As a concrete example for the above claim, lets consider the BPS STU-example of section 2 and,
for obtaining the black hole entropy as a Legendre transformation from \emph{some free energy}, keep (\ref{eq7}) unused. This way, from (\ref{eq1}-\ref{eq6}) one obtains,
\bea \label{notq0} (\;\;C_{2,j} = 0\;\;\;;\;\;\; t_2^j = - \frac{p^j}{C_2}\;\;) \eea with the attractor value of $C_2$ undetermined at this stage. Moreover we define, 
\bea\label{hatfree} \hat{\mathcal{F}}(\mathcal{D},C_2)\;\equiv\;-\pi\;\Im\;[\;C^{2} F(X)\;]_{|_{(\ref{notq0})}} =  -\pi\;\Im\;[\;F(C X)\;]_{|_{(\ref{notq0})}} \;\;\;\eea
which according to the relation (\ref{ff}) yields,
\bea\label{hatfreetop}  \hat{\mathcal{F}}( \mathcal{D} , C_2 )\;=\;2\; \Re\;[\;F_{top}(t^i,g_{top})\;]_{|_{(\ref{notq0})}} \;=\;2\; \Re\;[\;F_{top}(t^j,g_{top})\;]_{|_{(g_{top}\;=\pm \frac{4 \pi}{C_2}\;,\;t^j\;=\;\frac{p^j}{\mathbf{i}C_2})}}.\eea
From (\ref{hatfree}), (\ref{notq0}) and (\ref{STU}) one gets,
\bea\label{hatf} \hat{\mathcal{F}}( \mathcal{D}, C_2 ) = \pi \frac{\mathcal{D}}{C_2}\;\;\;.\eea
Now lets perform a Legendre transformation on $\hat{\mathcal{F}}$ over the variable $C_2$ and call it $\mathcal{L}_{C_2}[\hat{\mathcal{F}}]$,
\bea\label{Lf} \mathcal{L}_{C_2}[\hat{\mathcal{F}}]\;\equiv\;\hat{\mathcal{F}}( \mathcal{D} , C_2 )\;-\;C_2\;\frac{\partial \hat{\mathcal{F}}( \mathcal{D} , C_2 )}{\partial C_2}\;\;\;.\eea 
Using the result (\ref{hatf}) one gets,
\bea\label{RHS} \mathcal{L}_{C_2}[\hat{\mathcal{F}}]\;=\;2 \pi \frac{\mathcal{D}}{C_2}\;\;\;. \eea
It is turn to use (\ref{eq7}). Substituting the results of (\ref{notq0}) into (\ref{eq7}) gives us
the attractor value of $C_2$ which according to (\ref{susyAt}) is given by,
\bea\label{c2} C^{*}_2\;=\;\pm \sqrt{\frac{\mathcal{D}}{q_0}}\;\;\;.\eea
Now comparing the value of entropy as given by (\ref{entropy}) with (\ref{RHS}) and (\ref{c2}) one deduces,
\bea\label{deduc} S_{BH}\;=\;\mathcal{L}_{C_{2}^{*}} [\hat{\mathcal{F}}]\;\equiv\;\mathcal{L}_{C_2}[\hat{\mathcal{F}}]_{|_{C_2^{*}}}\;\;\;\;.\eea
This suggests the identification of our BPS black hole free energy with $\hat{\mathcal{F}}(\mathcal{D},C_2)$. Correspondingly the \emph{reduced mixed ensemble} of this BPS black hole is proposed as,
\bea\label{reduensemble} Z_{BH}(\mathcal{D},C) = \rm{e}^{\hat{\mathcal{F}}(\mathcal{D},C_2)} \;\equiv\; \sum_{N} \hat{\Omega}(N\mathcal{D})\; e^{-N\;C_2} \eea
with $N$ standing for the charge corresponding to $C_2$,
\bea\label{N}N\;\equiv\;-\;\frac{\partial \hat{\mathcal{F}}}{\partial C_2}\;_{|_{C_2^{*}}}\;=\;\pi q_0\eea
and $\hat{\Omega}$ being the same as OSV degeneracy of states, 
\bea\label{omegahat} \hat{\Omega}(\pi q_0 \mathcal{D})\;=\;\Omega(p,q)\;\;\;.\eea
Now according to (\ref{reduensemble}) and (\ref{hatfreetop}) one recovers the OSV relation in the form,
\bea\label{soOSV}Z_{BH}(\mathcal{D},C) = |Z_{top}(t^i,g_{top})|^2_{|_{(g_{top}\;=\pm \frac{4 \pi \mathbf{i}}{C}\;,\;t^j\;=\;\frac{p^j}{C})}}\;\;\;.\eea
Finally in order of relaxing the gauge fixing condition $X^0=1$, we define \bea\label{fi0}\phi^0 \equiv \pi \Im\;(CX^0)\eea and then using the map (\ref{map}) the generalization of the relation (\ref{soOSV}) is given by,
\bea\label{generali} Z_{BH}(D(p),\phi^0) \; =\; \rm{e}^{\frac{\pi^2 D(p)}{\phi^0}}\;\equiv \; \sum_{\pi\hat{q}_0} \hat{\Omega} (\pi D(p)\hat{q}_0)\; \rm{e}^{- \hat{q}_0\;\phi^0} \;=\; |Z_{top}(t^j,g_{top})|^2_{|_{(g_{top}\;=\pm \frac{4 \pi^2 }{\phi^0}\;,\;t^j\;=\;\frac{\pi p^j}{\mathbf{i} \phi^0})}} \eea
which is the case for all the BPS black holes with vanishing $D6$-brane charge in the supergravity limit.\\Here to highlight the equivalence of the ensemble (\ref{generali}) with the OSV ensemble (\ref{BPSensemble}), it is worth to explicitly calculate the OSV free energy (\ref{free}) for the BPS black hole carrying charges $(p^i,q_i,q_0)$ in the supergravity defined by (\ref{prepoten0}). It is evaluated as,
\bea\label{OSVfree0}\mathcal{F}(p^I,\phi_I) = \frac{\pi^2 D}{\phi^0} - \frac{3 D_{ab}\phi^a\phi^b}{\phi^0}\eea
which, since $C_2=\frac{\phi^0}{\pi}$ in the gauge $X^0=1$, differs from (\ref{hatf}) by the second term at the right hand side of (\ref{OSVfree0}). However (\ref{hatf}) and (\ref{OSVfree0}), although presenting different functions, define \emph{equivalent} free energies for the black hole under consideration, as seen by the equalities,
$$\mathcal{L}_{\{\phi^I\}}[\mathcal{F}]\;=\;\mathcal{L}_{C_2}[\hat{\mathcal{F}}]\;\;\;\;;\;\;\; \phi^0_{|_{q_I = - \frac{\partial \mathcal{F}}{\partial \phi^I}}}\;=\;\phi^0_{|_{\hat{q}_0 = - \frac{\partial \hat{\mathcal{F}}}{\partial \phi^0}}}$$
Although compared to the OSV relations (\ref{TSBH}) and (\ref{vari}), the relation (\ref{generali}) is more restricted since it does not incorporate neither the higher-genus terms nor the $D6$-brane charge, the \emph{procedure} which led to the ensemble (\ref{generali}) is of importance for us. We shall follow the same approach for the case of non-supersymmetric black holes in the next subsection.
%%%%%%%%%%%%%%%%%%%%%%%%%%%%%%%%%%%%%%%%%%%%%%%%%%%%%%%%%%%%%%%%%%%%%%%%%%%%%%%%%%%%%
\subsection{\underline{Partition function of the non-supersymmetric attractors}}
To find the partition function of the non-susy attractors in the supergravity limit, we begin with the non-susy sector of the STU model with $(p^0,q_i)=0$, as discussed in section 2, and after deriving the partition function, map the result to the general case of (\ref{prepoten0}) with nonvanishing $q_i$, via (\ref{map}).\\ The procedure is similar to what was done in the previous subsection for the susy sector.
If we keep the equation (\ref{eq7}) unsolved, the equations (\ref{eq1}-\ref{eq6}) together, for the non-susy sector, imply,
\bea \label{nsnotq0} (\;\;C_{2,j} = \frac{C_2}{t_2^j}\;\;\;;\;\;\; t_2^j = - \frac{p^j}{2\;C_2}\;\;) \eea 
Therefore defining, 
\bea\label{nshatfree} \hat{\mathcal{F}_{ns}}(\mathcal{D},C_2)\;\equiv\;-\pi\;\Im\;[\;C^{2} F(X)\;]_{|_{(\ref{nsnotq0})}} = -\pi\;\Im\;[\;F(C X)\;]_{|_{(\ref{nsnotq0})}} \;\;\;\eea

From (\ref{nshatfree}), (\ref{nsnotq0}) and (\ref{STU}) one gets,
\bea\label{nshatf} \hat{\mathcal{F}_{ns}}( \mathcal{D}, C_2 ) = \pi \frac{\mathcal{D}}{8\;C_2}\;\;\;.\eea
We perform a Legendre transformation on $\hat{\mathcal{F}_{ns}}$ over the variable $C_2$ and call it $\mathcal{L}_{C_2}[\hat{\mathcal{F}_{ns}}]$,
\bea\label{nsLf} \mathcal{L}_{C_2}[\hat{\mathcal{F}_{ns}}]\;\equiv\;\hat{\mathcal{F}_{ns}} ( \mathcal{D} , C_2 )\;-\;C_2\;\frac{\partial \hat{\mathcal{F}_{ns}}( \mathcal{D} , C_2 )}{\partial C_2}\;\;\;\eea 
which with (\ref{nshatf}) equals,
\bea\label{nsRHS} \mathcal{L}_{C_2}[\hat{\mathcal{F}_{ns}}]\;= \pi \frac{\mathcal{D}}{4\;C_2}\;\;\;. \eea
Further, lets compute the attractor value of $\mathcal{L}_{C_2}[\hat{\mathcal{F}_{ns}}]$ and compare with $S_{BH}$ as given by (\ref{entropy}). Substituting the results of (\ref{nsnotq0}) into (\ref{eq7}), the attractor value of $C_2$ is found to be,
\bea\label{nsc2} C^{*}_2\;=\;\pm \frac{1}{2} \sqrt{-\frac{\mathcal{D}}{q_0}}\;\;\;\eea
so that,
\bea\label{nsLfat} \mathcal{L}_{C_2^{*}} [\hat{\mathcal{F}_{ns}}]\;=\;\pm \frac{\pi}{2}\;\sqrt{-q_0\mathcal{D}}\;\;\;. \eea 
Now comparing (\ref{nsLfat}) with (\ref{entropy}) one observes,
\bea\label{nsdeduc} S_{BH}\;=\;4\;\mathcal{L}_{C_{2}^{*}} [\hat{\mathcal{F}_{ns}}]\;\;\;\;.\eea
Thus defining,
\bea\label{newfree} \mathcal{F}_{ns.BH}(\mathcal{D},C_2)\;\equiv\;4\;\hat{\mathcal{F}_{ns}}(\mathcal{D},C_2)\;\;\;\;\eea
one obtains,
\bea\label{newnsdeduc} S_{ns.BH}\;=\;\mathcal{L}_{C_{2}^{*}} [\mathcal{F}_{ns.BH}]\;\;\;\;.\eea
which completes the procedure. Thus the mixed partition function of this non-susy black hole is proposed as,
\bea\label{nsreduensemble} Z_{ns.BH}(\mathcal{D},C) = \rm{e}^{\mathcal{F}_{ns.BH}(\mathcal{D},C_2)} \;\equiv\; \sum_{M} \rm{d}(M\mathcal{D})\; e^{-M\;C_2} \eea
with $M$ regarded as the charge corresponding to $C_2$,
\bea\label{nsN}M\;\equiv\;-\;\frac{\partial \mathcal{F}_{ns.BH}} {\partial C_2}\;_{|_{C_2^{*}}}\;=\;-2 \pi q_0\eea
and $\rm{d}(M\mathcal{D})$ as the black hole degeneracy of states carrying charges $(p^1,p^2,p^3,q_0)$.\\ 
Finally defining $\phi^0 \equiv \pi \Im\;(CX^0)$ to relax the gauge fixing condition $X^0 = 1$, and using the map (\ref{map}), we generalize the above partition function to the case of general non-supersymmetric black holefor non-supersymmetric black holes attractors in the supergravity limit as follows,
\bea\label{newgenerali} Z_{ns.BH}(D(p),\phi^0)  = \rm{e}^{\frac{\pi^2 D(p)}{2 \phi^0}}\;\equiv\; \sum_{-2\pi\hat{q}_0} \rm{d} (-2\pi D(p)\hat{q}_0)\; e^{2 \hat{q}_0 \phi^0}  \eea
%%%%%%%%%%%%%%%%%%%%%%%%%%%%%%%%%%%%%%%%%%%%%%%%%%%%%%%%%%%%%%%%%%%
\section{Open questions}
Here we briefly mention two problems which call for further investigations:\\
$1.\;$ The result for $\rm{d}(p,q)$ as derived from the partition function (\ref{result}) can be checked   with an independent counting of the black hole degeneracy of states in a given set up. However for non-supersymmetric black holes it seems pretty nontrivial to find a set up in which such a counting is possible.\\
$2.\;$ The partition function (\ref{result}) is restricted to the supergravity limit. It is interesting to look for the partition function of non-supersymmetric black holes beyond this limit. Again
compared to BPS black holes it is a more nontrivial job to do. For example the role played by the D-terms and the holomorphic anomaly is mainly unknown for the case of non-supersymmetric black holes.
%%%%%%%%%%%%%%%%%%%%%%%%%%%%%%%%%%%%%%%%%%%%%%%%%%%%%%%%%%%%%%%%%%%
\begin{center}
{\large {\bf Acknowledgement}}
\end{center}
This is our pleasure to thank M. Alishahiha, A. Dabholkar, K.T. Joseph and S. P. Trivedi for useful discussions. A. T. would like to thank the Department of Theoretical Physics at TIFR for warm hospitality. We are especially grateful to S. Minwalla  and C. Vafa for their insightful comments about the draft.
%%%%%%%%%%%%%%%%%%%%%%%%%%%%%%%%%%%%%%%%%%%%%%%%%%%%%%%%%%%%%%%%%%%

\end{document}